\documentclass[12pt]{iopart}
\usepackage{iopams}  

\usepackage{graphicx}
\usepackage{float}

\def\D{\Delta}

\def\a{\alpha}
\def\b{\beta}
\def\k{\kappa}
\def\W{\Omega}
\def\g{\gamma}

\def\d{\delta}
\def\p{^{\prime}}

\newcommand{\ket}[1]{\left| {#1} \right\rangle}

\def\be{\begin{equation}}
\def\ee{\end{equation}}
\def\bea{\begin{eqnarray}}
\def\eea{\end{eqnarray}}

\def\>{\right\rangle}
\def\<{\left\langle}

\begin{document}

\title{On the indistinguishability of Raman photons}

\author{Charles Santori$^1$, David Fattal$^2$, Kai-Mei C. Fu, Paul E. Barclay, and Raymond G. Beausoleil}
\address{Hewlett-Packard Laboratories, 1501 Page Mill Road, Palo Alto, CA 94304}
\ead{$^1$charles.santori@hp.com, $^2$david.fattal@hp.com}

%%%%%%%%%%%%%%%%%%%%%%%%%%%%%%%%%%%%%%%%%%%%%%%%%%%%%%%%%%%%%%%%%%%%%%%%%%%%%

\begin{abstract}
We provide a theoretical framework to study the effect of dephasing on the quantum indistinguishability of single photons emitted from a coherently driven cavity QED $\Lambda$-system.  We show that with a large excited-state detuning, the photon indistinguishability can be drastically improved provided that the fluctuation rate of the noise source affecting the excited state is fast compared with the photon emission rate.  In some cases a spectral filter is required to realize this improvement, but the cost in efficiency can be made small.
\end{abstract}


%03.67.Lx Quantum computation architectures and implementations
%33.50.Dq Fluorescence and phosphorescence spectra
%42.25.Kb Coherence
%42.50.Ar Photon statistics and coherence theory
%42.50.Ct Quantum description of interaction of light and matter; related experiments
%42.50.Dv Quantum state engineering and measurements
%42.50.Ex Optical implementations of quantum information processing and transfer in quantum optics
%42.50.Pq Cavity quantum electrodynamics; micromasers
%61.72.J- Point defects and defect clusters
%76.30.Mi Color centers and other defects
%78.67.-n Optical properties of low-dimensional, mesoscopic, and nanoscale materials and structures
%78.67.Hc Quantum dots

\maketitle

\section{Introduction}

Indistinguishable photons are required for many quantum computation schemes that utilize their peculiar interference properties~\cite{ref:knill2001seq,ref:childress2005ftq}. They reveal a true bosonic nature and for instance tend to coalesce or ``bunch'' when they cross at a beam-splitter~\cite{ref:hong1987mst}.  Indistinguishable photons are naturally produced during the spontaneous decay of atoms and ions~\cite{ref:beugnon2006qib,ref:maunz2007qip}. Solid-state atom-like systems can also emit indistinguishable photons, with examples including quantum dots~\cite{ref:santori2002ipf,ref:laurent2005isp,ref:bennett2005ied}, molecules~\cite{ref:kiraz2005ips} and semiconductor impurities \cite{ref:sanaka2009ipi}; diamond nitrogen-vacancy (NV) centers also appear promising~\cite{ref:tamarat2006ssc}. Solid-state systems are especially interesting for large-scale photonic networks incorporating many devices.  The temporal length of the emitted photons can be short due to the large dipole moment and correspondingly short spontaneous emission lifetime ($0.5-12\,\mathrm{ns}$), and the lifetime can be further shortened using integrated small mode volume optical cavities.  The main drawback of solid-state systems is that they can suffer from severe dephasing of the optical transitions caused by phonons~\cite{ref:besombes2001apb}, random charge fluctuations~\cite{ref:seufert2000sde} or other mechanisms~\cite{ref:ambrose1991fsa}.

One way to reduce the detrimental effect of dephasing on photon indistinguishability in two-level systems is to push towards shorter-lived transitions.  This approach reaches its limit when the photon lifetime becomes comparable to the time jitter that characterizes incoherently pumped transitions.  This then necessitates true resonant excitation, which has recently been demonstrated~\cite{ref:muller2007rfc} but limits experiments to cavity geometries that minimize pump laser scattering.  Alternatively, one can move to three-level systems that can be coherently excited at a frequency different from that of the photon emission.  Three-level $\Lambda$-type systems also potentially provide an interface between long-lived matter qubits and ``flying'' photonic qubits.  Schemes based on Raman transitions in three-level systems were proposed more than a decade ago for deterministic communication in quantum networks~\cite{ref:cirac1997qst}, and for probabilistic entanglement generation~\cite{ref:cabrillo1999ces} and teleportation~\cite{ref:bose1999pta}.  Some more recent papers have investigated the theoretical aspects of single-photon generation in three-level systems in greater detail, including the problem of generation and trapping of photons with arbitrary waveforms~\cite{ref:yao2005tcs,ref:fattal2006csp}, and the problem of how best to generate indistinguishable photons from non-identical systems~\cite{ref:cancellieri2009ogi}.  Experimentally, three-level schemes have now been used for single-photon generation in atoms~\cite{ref:kuhn2002dss}, and in trapped ions~\cite{ref:keller2004cgs} it was shown that photons with arbitrary waveforms could be generated.  Such experiments have not yet been performed in solid-state systems, but following a recent demonstration of tunable, spontaneous Raman fluorescence from a single quantum dot~\cite{ref:fernandez2009ots} the outlook is promising.

For solid-state implementations of the Raman scheme, a critical question is how the photon indistinguishability is affected by excited-state dephasing.  Ref.~\cite{ref:kiraz2004qss} investigated this question but only for the special case of zero detuning, and furthermore the model, based on optical Bloch equations, did not allow for finite correlation timescales in the dephasing process.  The results suggested that the three-level scheme is helpful for solving the time-jitter problem, but little or no improvement with respect to excited-state dephasing was indicated.  One might expect that for a large detuning, single photons generated in a Raman process will be far less sensitive to excited-state energy fluctuations.  In this paper, we analyze theoretically the dependence of photon indistinguishability on both the detuning and the correlation timescale of the dephasing process affecting the excited state.  We find that for a large detuning, the effects of excited-state dephasing can be reduced by orders of magnitude beyond the usual reduction that occurs due to shortened photon lifetime in a two-level system, but only if the memory timescale of the noise process is short compared with the photon emission lifetime.

\section{Three-Level Equations of Motion in the Schr\"{o}dinger Picture}

\begin{figure}[t]
    \centering
    \includegraphics[width=4in]{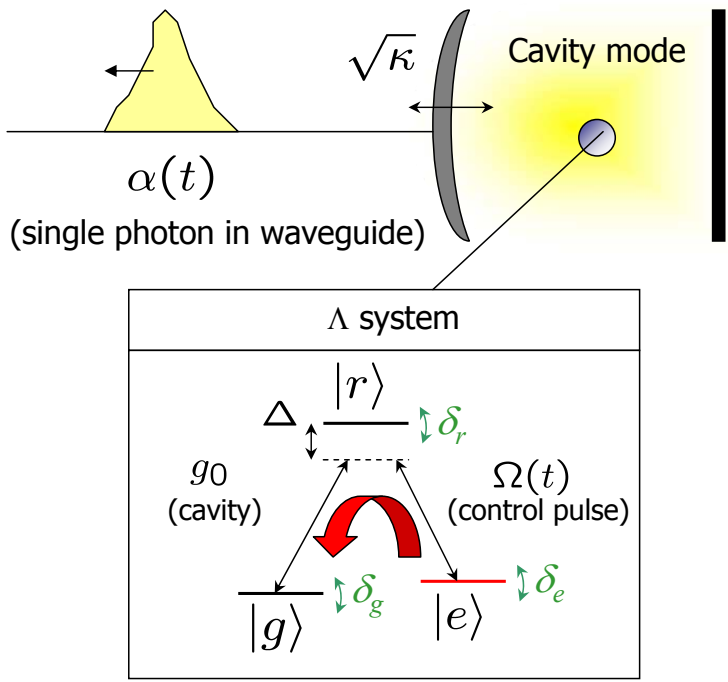}
    \caption{A $\Lambda$-system in a single-mode cavity coupled to a single-mode waveguide.}
    \label{fig:scheme}
\end{figure}
Consider a 3-level system as in Fig.~\ref{fig:scheme}, with two ground states $\ket{e}$ and $\ket{g}$ and an excited state $\ket{r}$ where the $e-r$ and $g-r$ transitions are optically allowed. The $g-r$ transition couples to an optical cavity mode with Rabi vacuum frequency $g_0$ and detuning $\D$, and the cavity mode itself is coupled to a single-mode waveguide and has total decay rate $\k$. The excited state $\ket{r}$ can decay by emitting a photon outside of the cavity with rate $\g$ (not desired but nevertheless present).  The system starts in level $\ket{e}$ and is exposed to a classical time-varying light beam which drives the $e-r$ transition with detuning $\D$ (so that the e-to-r-to-g transition is on two-photon resonance) and complex Rabi frequency $\W(t)$.  We neglect coupling of the laser to the $g-r$ transition and coupling of the cavity to the $e-r$ transition, good approximations only if we have either good polarization selection rules or an energy spacing between $\ket{e}$ and $\ket{g}$ that is much larger than $\D$ and $\kappa$.  In this calculation we include spontaneous decay only from state $\ket{r}$ to state $\ket{g}$, so that the system cannot be re-excited after emitting a photon.  This allows for a large simplification of the dynamics, since only states with zero or one photon occur.  Spontaneous decay back to the initial state $\ket{e}$ can, in fact, be an important decoherence process, and the effect on photon indistinguishability was analyzed in Ref.~\cite{ref:kiraz2004qss}.  Here we are interested mainly in the effect that pure dephasing of the excited state has on the emitted photons, so we set the $r \rightarrow e$ decay rate to zero.  While there are practical limitations, the $r \rightarrow e$ rate can be made small compared with the $r \rightarrow g$ rate either by having unequal strengths for the two optical transitions or by using a large Purcell enhancement of the $g-r$ transition.  Below, we shall consider the $r \rightarrow e$ decoherence process separately when discussing the application of this scheme to actual solid-state systems.

\subsection{Unperturbed Evolution of a Three-Level $\Lambda$ System}

First let us consider a deterministic trajectory of the system without dephasing processes.  A Hamiltonian describing the three-level system, cavity, and continuum modes outside of the cavity under the assumptions given above and under the rotating-wave approximation is,
\bea
\fl
H_0 &=& \sum_i \omega_i \sigma_{ii} + 
\frac{\W}{2} e^{i\omega_l t}\sigma_{er} + \frac{\W^\ast}{2} e^{-i\omega_l t} \sigma_{re} +
\omega_c c^{\dagger} c +
g_0 \sigma_{gr} c^{\dagger} + g_0^\ast \sigma_{rg} c + \nonumber \\
\fl
& & \sum_k \left( \omega_k a_k^\dagger a_k + h_k c \, a_k^{\dagger} + h_k^\ast c^{\dagger} a_k \right) + 
\sum_{k^\prime} \left( \omega_{k^{\prime}} b_{k^{\prime}}^\dagger b_{k^{\prime}} + g_{k^{\prime}} \sigma_{gr} b_{k^{\prime}}^{\dagger} + g_{k^{\prime}}^\ast \sigma_{rg} b_{k^{\prime}} \right) \, ,
\eea
where $\sigma_{ij} = |i\rangle \langle j|$ for $i,j = \{e, r, g\}$, and $c$, $a_{k}$ and $b_{k^{\prime}}$ are photon annihilation operators for a single cavity mode, the continuum modes coupled to this cavity mode, and the continuum modes coupled to the $g-r$ transition, respectively. The constants $\omega_i$ are the frequencies of the three levels, $\omega_c$ is the cavity resonance frequency, $\omega_l$ is the frequency of the excitation laser, $\omega_k$ and $\omega_{k^{\prime}}$ are the continuum mode frequencies, and $h_k$ and $g_{k^{\prime}}$ are the coupling constants to the continuum modes.  We work in a rotating frame by writing the state of the system (assuming we begin in state $|g\rangle$ so that only one photon can be emitted) as,
\bea
|\psi (t) \rangle &=& e(t) e^{-i\omega_e t} |e\rangle +
                    r(t) e^{-i(\omega_r-\Delta)t} |r\rangle +
                    g(t) e^{-i(\omega_g+\omega_c-\delta_c)t} c^{\dagger} |g\rangle + \nonumber \\
& &    \sum_k \alpha_k(t) e^{-i(\omega_g+\omega_k)t} a_k^{\dagger} |g\rangle +
       \sum_{k^{\prime}} \alpha^{\prime}_{k^{\prime}}(t) e^{-i(\omega_g+\omega_{k^{\prime}})t} b_{k^{\prime}}^{\dagger} |g\rangle \, ,
\eea
where $e(t)$ and $r(t)$ are the amplitudes of the system in states $\ket{e}$ or $\ket{r}$ with no photon in the cavity, and $g(t)$ is the amplitude of the system in state $\ket{g}$ with one photon in the cavity.  We have assumed that direct spontaneous emission from level $r$ and photon escape from the cavity involve orthogonal sets of radiation modes, and $\alpha_k(t)$ and $\alpha_{k^{\prime}}(t)$ are the amplitudes for photon emission into these modes.  The detuning constants for the laser and cavity are $\Delta = \omega_r - \omega_e - \omega_l$ and $\delta_c = \omega_c - \omega_r + \omega_g + \Delta$, respectively.  In the calculation below we set $\delta_c = 0$, though this parameter could be adjusted to compensate for a.c. Stark shifts if desired.  By writing $\frac{d}{dt}|\psi(t)\rangle = -i H_0(t) |\psi(t)\rangle$ and applying the Weisskopf-Wigner approximation to remove the continuum modes from the dynamics, we find the unperturbed equations of motion,
\bea \dot{e}(t) & = & -i \frac{\W}{2} r(t) \label{eq:edot} \, ,\\
\dot{r}(t) & = & -\left(i \D + \frac{\g}{2}\right) r(t) -i \frac{\W^{\ast}}{2} e(t) -i g_0^{\ast} g(t) \label{eq:rdot} \, ,\\
\dot{g}(t) & = & -i g_0 r(t) -\left(i \delta_c + \frac{\k}{2} \right) g(t) \label{eq:gdot} \, ,\eea
which are similar to those used in Refs.~\cite{ref:yao2005tcs,ref:fattal2006csp}.  These equations include decay rates $\kappa$ and $\gamma$ for the cavity and level $r$ that can be related to the $h_k$ and $g_{k^{\prime}}$ coupling coefficients, respectively, combined with the corresponding densities of states for the continuum modes.  Following the cavity input-output formalism, we can also define a temporal envelope $\alpha(t)$ of the photon emitted into the waveguide as the Fourier transform of $\alpha_k(t \rightarrow \infty)$ with respect to $\omega_k$, including only the subset of modes that belong to the waveguide.  This temporal envelope is simply $\a(t) = \sqrt{\kappa_{\mathrm{wg}}} g(t)$, where $\kappa_{\mathrm{wg}} < \kappa$ is the cavity decay rate into the waveguide modes only.  The coherence properties of $\a(t)$ are thus the same as those of $g(t)$.  The efficiency to emit a photon into the waveguide is $\eta = \int dt \, |\a(t)|^2$.

For the unperturbed dynamics we shall consider only the adiabatic limits $\dot{g} \ll \kappa g$ and $\dot{r} \ll \D r$ for which the unperturbed dynamics follow simple analytical solutions,
\bea
e(t) &=& \exp \left[ \frac{i}{4\D\p} \int^t_{-\infty} dt\p |\W(t\p)|^2 \right] \, , \label{eq:e_approx}\\
r(t) &=& - \frac{\Omega^{\ast}(t)}{2 \D\p} e(t) \, , \label{eq:r_approx}\\
g(t) &=& -\frac{2i g_0}{\kappa} r(t) \, . \label{eq:g_approx}
\eea
Here, we have defined $\D\p = \D -i\gamma_p/2$ as the complex detuning and $\gamma_p = \gamma + 4|g_0|^2/\kappa$ as the cavity-enhanced spontaneous emission rate from level $r$.  This approximate solution holds if $\{ |\Omega|, \, \gamma_p \} \ll |\Delta|$, $\left|\frac{\Omega \, g_0^{\star}}{\Delta} \right | \ll \kappa$, $\frac{|\Omega|^2}{4 \Delta^2} \gamma_p \ll \gamma$, and  $\left| \frac{|\Omega|^2}{4\Delta} + \delta_c \right| \ll \kappa$.

\subsection{Evolution and Indistinguishability in a Randomly Fluctuating Environment}

For the deterministic Eqs.~\ref{eq:edot}-\ref{eq:gdot}, $\a(t)$ is always the same, and consecutive photons emitted by the system are perfectly indistinguishable.  Now, we wish to describe the real system which is affected by random fluctuations in its environment.  Various treatments have been developed to include decoherence effects in the context of quantum information, the most commonly used being the optical Bloch equations~\cite{ref:cohentannoudji1992api} and quantum jump approaches~\cite{ref:plenio1998qja}.  While a quantum jump approach may be preferable in terms of simplicity and physical insight, both approaches are typically used to describe interactions with a reservoir having infinitely short correlation timescales.  In solid-state systems, dephasing processes exist with a wide range of memory timescales spanning at least $10^{-15} - 10^{3} \, \mathrm{s}$.  To describe pure dephasing processes with finite memory timescales, we follow an alternative approach that uses randomly fluctuating energy shifts to represent these processes~\cite{ref:anderson1954mmn,ref:kubo1954nst,ref:daffer2004dcc,ref:ban2006dqi}.  The time-dependent energy shifts of the system levels are denoted by $\d_e(t)$, $\d_r(t)$ and $\d_g(t)$, or $\vec{\d}$ for short, with correlation properties to be defined below.  If the unperturbed equations of motion are written as $\frac{d}{dt} \vec{x} = M_0(t) \vec{x}$, where $\vec{x}$ is a vector containing $e$, $r$, and $g$, and $M_0(t)$ is a $3 \times 3$ matrix based on Eqs.~\ref{eq:edot}-\ref{eq:gdot}, then the perturbed equations of motion are,
\begin{equation}
\frac{d}{dt} \tilde{\vec{x}} = M_0(t) \tilde{\vec{x}} - i \,\mathrm{diag} ({\vec{\delta}}(t)) \, \tilde{\vec{x}} \, , \label{eq:xtildedot}
\end{equation}
where $\tilde{x_i}$ are the perturbed amplitudes.  Changing variables, let us define,
\begin{equation}
\tilde{x_i}(t) = x_i(t)\exp(-i\phi_i(t)) \, , \label{eq:phidef}
\end{equation}
where the $\phi$'s are arbitrary complex functions of time.  We substitute this into Eq.~\ref{eq:xtildedot} (using Eqs.~\ref{eq:edot}-\ref{eq:gdot} for $M_0(t)$) and linearize to first order in $\vec{\phi}$ assuming that the fluctuations are small.  The linearized equations can be written as $\frac{d}{dt} \vec{\phi} - \vec{\d} = M(t) \vec{\phi}$, where,
\be M(t) = \left(\begin{array}{ccc}
i\frac{\W}{2}\frac{r}{e} & -i\frac{\W}{2}\frac{r}{e} & 0 \\
-i\frac{\W^{\ast}}{2}\frac{e}{r} & i\frac{\W^{\ast}}{2}\frac{e}{r} + i g_0^{\ast} \frac{g}{r} & -i g_0^{\ast} \frac{g}{r} \\
 0 & -i g_0 \frac{r}{g} & i g_0 \frac{r}{g}
\end{array}\right) \label{eq:phi_dynamics} \, . \ee

For the noise-induced dephasing, here we shall consider only dephasing of the excited state, setting $\d_g(t) = \d_e(t) = 0$.  Let us start with the simple case where $\kappa$ is large compared with $g_0$, $\Delta$, and the fluctuation rate of the noise source.  In this limit the cavity serves only to enhance the effective spontaneous emission rate from level $r$, and otherwise the cavity does not enter into the problem.  Using $|\dot{\phi_g}| \ll \kappa |\phi_g|$ and substituting Eqs.~\ref{eq:e_approx}-\ref{eq:g_approx} into Eq.~\ref{eq:phi_dynamics} with $\vec{\phi}(-\infty) = 0$ we find,
\bea
\phi_e(t) &=& \frac{1}{4{\D\p}^2} \int_{-\infty}^t dt\p |\W(t\p)|^2 \bar{\delta}(t\p) \, , \label{eq:phie_approx}\\
\phi_g(t) &=& \phi_r(t) = \phi_e(t) + \frac{\bar{\delta}(t)}{i\D\p} \, , \label{eq:phig_approx}\\
\bar{\delta}(t) &=& i \D\p \int_{-\infty}^t dt\p \, e^{-i \D\p (t-t\p)} \d_r(t\p) \, \label{eq:deltabar} .
\eea
The phase $\phi_g(t)$ that is imprinted directly onto the emitted photon has two parts.  The first term $\phi_e(t)$ results from the accumulated phase error transferred to state $\ket{e}$ due to the fluctuating detuning combined with the a.c. Stark shift.  This term is dominant if the noise source fluctuates slowly, in which case $\bar{\delta}(t) \approx  \delta_r(t)$.  The second term represents a phase added directly to state $\ket{r}$ through its own energy fluctuation.

The indistinguishability of two photons emitted by identical devices as measured in a Hong-Ou-Mandel-type experiment~\cite{ref:hong1987mst} is,
\be
F = \frac{\< \left| \int dt \, \a_1(t) \a^\ast_2(t) \right|^2 \>}{\left( \int dt \, \< |\alpha(t)|^2 \> \right)^2} \, , \label{eq:Vdef}
\ee
where $\< \>$ denotes an expectation value over all possible functions $\delta_r(t)$, and $\alpha_1(t)$ and $\alpha_2(t)$ are the temporal envelopes of the two photons.  It is assumed that each photon is generated through an independent but identical random process.  From here on, let us consider the special case of a classical field that is turned on at $t=0$ and then held constant, $\W(t) = \W \, \theta(t)$.  In this case the unperturbed, approximate solution from Eqs.~\ref{eq:e_approx}-\ref{eq:g_approx} is a photon envelope with an abrupt rise at $t=0$ followed by an exponential decay, $\alpha(t) \propto e^{(-\xi/2 -i\zeta) t - i\phi_g(t)}$ where $\zeta \approx -\frac{|\W|^2}{4\D}$ and $\xi \approx \frac{|\W|^2}{4\D^2} \gamma_p$.  If the noise perturbation is small, to second order in $\phi_g$ the indistinguishability can be approximated as,
\bea
F & \approx & 1 + 2 \xi^2 \int_0^{\infty} ds \int_0^{\infty} dt \, e^{-\xi(s+t)} \mathrm{Re} \<\phi_g(s) \phi_g^\ast(t) \> \nonumber \\
& & - 2\xi \int_0^{\infty} dt \, e^{-\xi t} \< \phi_g(t) \phi_g^\ast(t) \> \, ,
\label{eq:Vforsmallphi}
\eea
where $\phi_g(t)$ may be complex.

\section{Finite-Memory Dephasing Processes}

Next, let us consider a specific noise process with zero mean and two-time correlation function,
\be
\< \d_r(t) \, \d_r(t+\tau) \> = \rho(\tau) = \sigma^2 e^{-\b |\tau|} \, ,
\label{eq:noisedef}
\ee
where $\sigma$ is the noise amplitude and $\b$ is the fluctuation rate (inverse correlation time).  This correlation function was also used in Refs.~\cite{ref:daffer2004dcc,ref:ban2006dqi} and can describe, for example, fluctuation of nearby traps that jump between two charge states, a common source of spectral diffusion of the optical transitions in semiconductor quantum dots and nitrogen-vacancy centers in diamond.  Suppose we have an ensemble of fluctuating traps that shift the energy (through the dc Stark effect, for example) of level $r$ according to,
\be
E_r(t) = \sum_n c_n q_n(t) \, ,
\ee
where $c_n$ are constants that depend on geometry, and $q_n = \{0, 1\}$ are random variables corresponding to the charge of each trap. The $q_n$ fluctuate independently according to,
\be
\dot{p_i} = \sum_j r_{ij} p_j \, ,
\ee
where $p_i$ is the probability that $q = i$, and $r_{ij}$ are the transition rates.  From this we can define a zero-mean random process $\delta_r(t) = E_r(t) - \langle E_r \rangle$ with correlation function given by Eq.~\ref{eq:noisedef} where,
\bea
\sigma^2 &=& \sum_n c_n^2 r_{10} r_{01} / (r_{01} + r_{10})^2 \, , \\
\beta &=& r_{01} + r_{10} \, .
\eea
The number of traps involved in the dynamics will affect higher-order correlation functions but will not change the form of the second-order correlations needed for the small-signal analysis presented here.  In experiments, the fluctuation timescale $1/\beta$ can vary of a wide range.  For diamond N-V centers probed occasionally by resonant excitation the timescale for spectral jumps can be many seconds but becomes much faster under non-resonant optical excitation at shorter wavelengths.  Of course, other pure dephasing processes may involve correlation functions with different functional forms.  For example, pure dephasing by phonons involves state-dependent scattering of an incident phonon to a new momentum state.  The spectral density for this process is determined by the phonon density of states combined with Bose-Einstein occupation probabilities~\cite{ref:roszak2005pid}.  The effective fluctuation rate is $\sim kT / \hbar$, but the form of the correlation function differs from that of Eq.~\ref{eq:noisedef}, which corresponds to a Lorentzian power spectrum.  To simplify our discussion, here we shall focus on the simple process described by Eq.~\ref{eq:noisedef}.  If a more accurate description is required, Eqs.~\ref{eq:Hdef}-\ref{eq:F_approx_freqdomain} presented below can be used to estimate the photon indistinguishability for arbitrary correlation functions.  Pure dephasing in the $\delta$-correlated limit, as used in the standard master equation or quantum jump approaches, can be obtained from Eq.~\ref{eq:noisedef} by taking the limit $\beta \rightarrow \infty$ with $\sigma^2/\beta \equiv 1/T_2$ held constant.

The photon indistinguishability under the process described by Eq.~\ref{eq:noisedef} can be calculated using Eqs.~\ref{eq:phie_approx}-\ref{eq:deltabar} and \ref{eq:Vforsmallphi} under the assumption that $\sigma$ is small so that $\phi_g$ is also small.  In the limit $|\Delta| \gg \{ |\Omega|, \gamma_p, \xi \}$, an approximate solution is,
\be
1 - F \approx \frac{2\sigma^2}{\gamma_p^2 (1 + \frac{\b}{\xi})} + 
\frac{2\sigma^2}{\D^2 + \b^2} \frac{\gamma_p + 2\b}{\gamma_p} \, .
\label{eq:indist_analytical}
\ee
The first and second terms in this expression correspond directly to $\phi_e(t)$ and $\bar{\delta}/i\Delta^\prime$, respectively, in Eq.~\ref{eq:phig_approx}.  The cross-correlation term was not included because, for large $\Delta$, it is always small compared with at least one of the terms in Eq.~\ref{eq:indist_analytical}.  For $\beta$ small compared with the photon emission rate, the first term, corresponding to the accumulated phase error on state $|e\rangle$, is dominant.  For large $\beta$, the first term becomes small since the noise is mostly averaged out in the integral of Eq.~\ref{eq:phie_approx} (in the frequency domain this acts as a low-pass filter).  This leaves the second term, which passes a larger noise bandwidth, to make the dominant contribution.
\begin{figure}[t]
    \centering
    \includegraphics[width=4.5in]{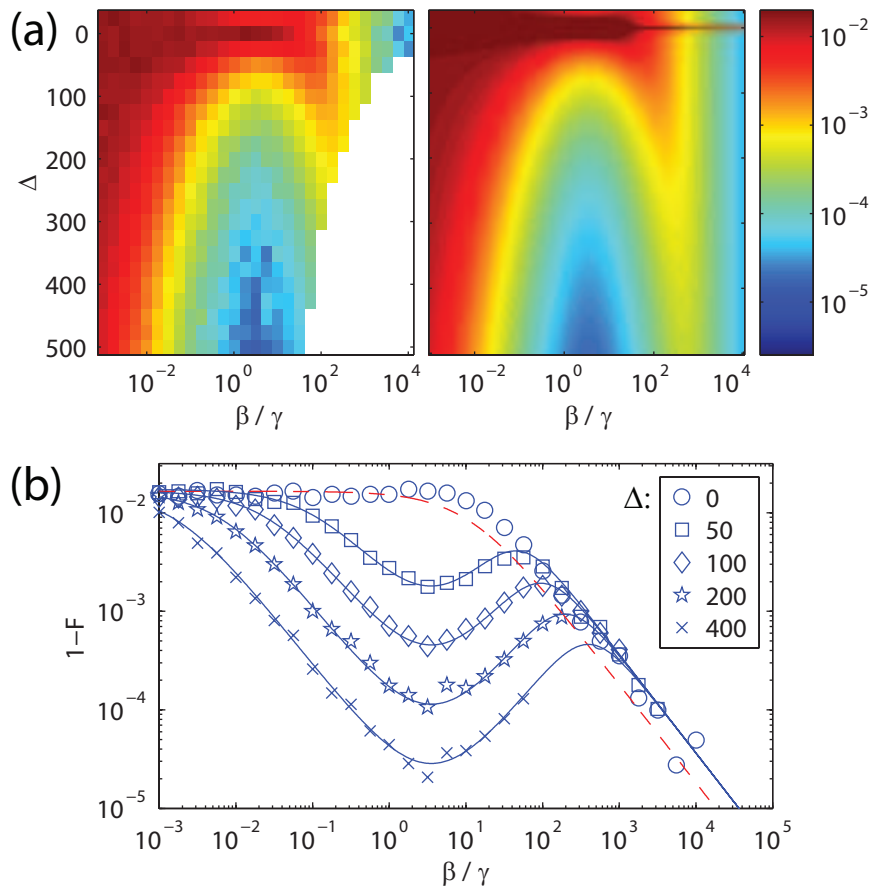}
    \caption{(a) Deviation from perfect photon indistinguishability ($1-F$) for the three-level system in the large-$\kappa$ regime with $g=50$, $\kappa=1000$, $\gamma=1$, $\sigma=1$ and $\Omega=10$ plotted on a logarithmic intensity scale as a function of detuning $\Delta$ and the normalized noise fluctuation rate ($\beta/\gamma$).  The plot on the left was obtained by the Monte-Carlo method.  The white region at lower-right was skipped due to excessively long computation times.  The plot on the right was obtained from Eq.\ref{eq:indist_analytical}. (b) Log-log plots of $1-F$ vs. $\beta / \gamma$ for various values of $\Delta$ as indicated.  The points (various symbols) are from the Monte-Carlo simulation, and the blue, solid curves are from Eq.\ref{eq:indist_analytical}.  The result for an instantaneously excited two-level system (Eq.~\ref{eq:V_twolevel_approx}) is also shown (red, dashed).}
    \label{fig:simulations}
\end{figure}
To test the accuracy of this approximate result we performed Monte-Carlo simulations in which the original Eqs.~\ref{eq:edot}-\ref{eq:gdot} are numerically integrated using a Gaussian-distributed noise source $\delta_r(t)$ with the required temporal correlations obtained from a pseudo-random number generator combined with a first-order finite-difference equation.  The number of trials was set so that the random errors in $1-F$ were usually less than 10\% of the mean value.  Figure~\ref{fig:simulations} shows calculated results in the $\kappa \gg \Delta$ regime (see figure caption for parameter values), and the results are in good agreement with Eq.~\ref{eq:indist_analytical} away from $\Delta=0$.

It is useful also to compare the photon indistinguishability predicted for the Raman scheme with that of an equivalent two-level system.  For a two-level system that is rapidly initialized into its excited state, if we neglect time jitter associated with the excitation processes, in the limits $g \ll \kappa$ and $\sigma\rightarrow 0$ the photon indistinguishability in the small-noise limit is,
\be
(1 - F)_{\mathrm{two-level}} \approx \frac{2\sigma^2}{\gamma_p (\gamma_p + \b)} \, .
\label{eq:V_twolevel_approx}
\ee
This is plotted alongside the results from the three-level system in Fig.~\ref{fig:simulations}(b) (red-dashed curve).  The results show that the Raman scheme can provide a substantial improvement over the two-level case only when the fluctuation rate $\beta$ of the noise source falls within the window $\xi < \beta < \Delta$.  When this condition is satisfied, an averaging effect occurs over the length of the photon which improves the indistinguishability.

\section{Indistinguishability Enhancement Using Spectral Filtering}

The first and second terms in Eq.~\ref{eq:indist_analytical} can also be understood in terms of a spectral function describing the transfer of noise from $\delta_r(t)$ to the emitted photon.  Let us define,
\be
f(\omega+i\eta/2) = \frac{1}{\sqrt{2\pi}} \int_0^\infty dt \, e^{\left(i \omega - \eta/2\right)t} \phi_g(t) \, ,
\ee
which can be written in terms of $\delta_r(t)$ as,
\be
f(\omega+i\eta/2) = - \left( \left[ M + \left(i\omega-\eta/2 \right)I \right]^{-1} \right)_{32} \,
\frac{1}{\sqrt{2\pi}}
\int_0^\infty dt \, e^{\left(i\omega-\eta/2\right) t} \delta_r(t) \, .
\ee
The energy spectral density $\langle |f(\omega+i\xi/2)|^2\rangle$ of noise transferred to the emitted photon is then proportional to the product of a transfer function,
\be
H(\omega, \eta) = \left| \left( \left[ M + \left(i\omega-\eta/2 \right)I \right]^{-1} \right)_{32} \right|^2 \, ,
\label{eq:Hdef}
\ee
and a weighted power spectral density of $\delta_r(t)$,
\be
S_r(\omega, \eta) =
\int_{-\infty}^\infty d\tau \rho(\tau) e^{i\omega\tau - \frac{\eta}{2}|\tau|} =
\frac{\sigma^2(\eta + 2\beta)}{\omega^2 + \left(\eta/2 + \beta \right)^2} \, ,
\ee
evaluated for $\eta = \xi$.  The photon indistinguishability in the small-$\phi_g$ limit given by Eq.~\ref{eq:Vforsmallphi} can be expressed in terms of these quantities as,
\be
F = 1 +  \xi H(0, 2\xi) S_r(0, 2\xi) - 2 \int \frac{d\omega}{2\pi} H(\omega,\xi) S_r(\omega, \xi) \, .
\label{eq:F_approx_freqdomain}
\ee

For the large-$\kappa$ approximation made above, $H(\omega,\xi)$ receives significant contributions from two poles at $\omega \approx \{-i\xi/2, \, \Delta - i\gamma_p/2 \}$.  The first pole corresponds to broadening of the Raman line itself, while the second corresponds to spontaneous emission at the natural frequency of the $r \rightarrow g$ transition.  For $\Delta \gg \gamma$, we claim that this second contribution can be filtered out with little effect on the photon collection efficiency.  We then expect only the first term in Eq.~\ref{eq:indist_analytical} to survive, giving,
\be
(1 - F)_{\mathrm{filter}} \approx \frac{2\sigma^2}{\gamma_p^2 (1 + \frac{\b}{\xi})} \, .
\label{eq:indist_analytical_filter}
\ee
Thus, with spectral filtering, the only remaining requirement to improve the indistinguishability over the two-level case is $\beta \gg \xi$.

\begin{figure}[t]
    \centering
    \includegraphics[width=4in]{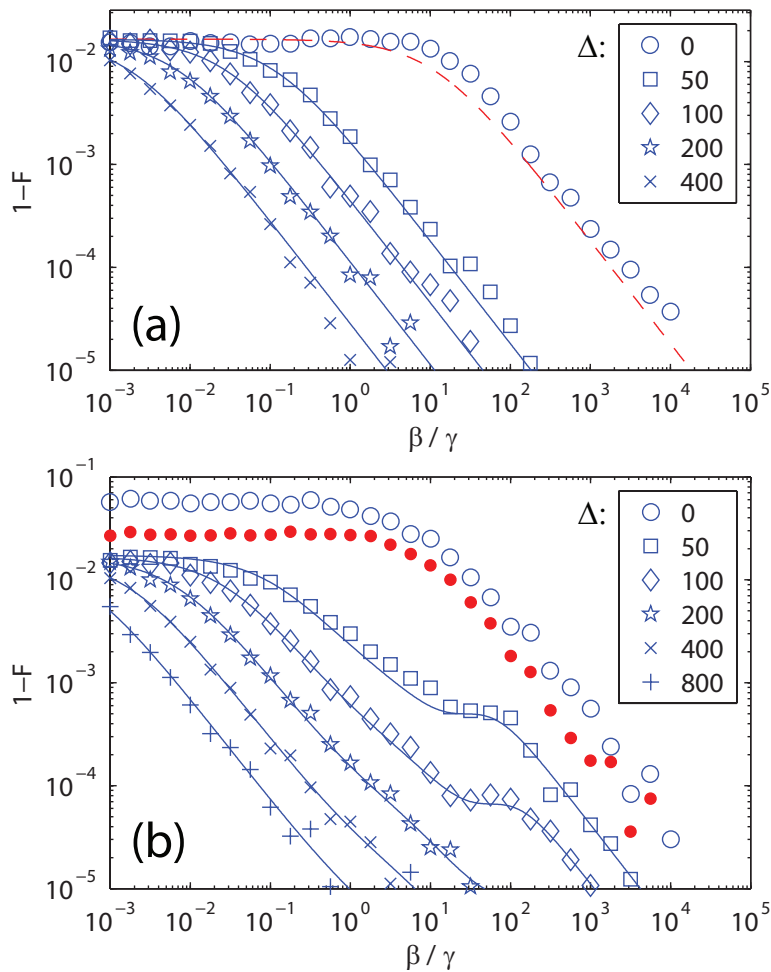}
    \caption{ Effect of two spectral filtering methods on photon indistinguishability: (a) An external spectral filter is centered on the Raman scattering wavelength to reject spontaneous emission at the natural transition frequency.  Parameters: $g=50$, $\kappa=1000$, $\gamma=1$, $\sigma=1$ and $\Omega=10$.  A simple Lorentzian filter was used with FWHM linewidth set to $20 \xi$ (see text).  The points (various symbols) are from Monte Carlo simulations, and the solid blue curves use only the first term of Eq.~\ref{eq:indist_analytical}.  The result for an instantaneously excited two-level system (Eq.~\ref{eq:V_twolevel_approx}) is also shown (red, dashed).  (b) The cavity itself serves as the external filter.  Parameters: $g=5$, $\kappa=10$, $\gamma=1$, $\sigma=1$ and $\Omega=10$.  The points are from Monte Carlo simulations, and the solid blue curves use Eq.~\ref{eq:indist_goodcavity}.  For the equivalent two-level system (filled red circles) a Monte Carlo simulation was used since Eq.~\ref{eq:V_twolevel_approx} is not accurate for $g \sim \kappa$.}
\label{fig:filter}
\end{figure}
To demonstrate that an external spectral filter can indeed improve the indistinguishability with little reduction in efficiency, Monte Carlo simulations of Eqs.~\ref{eq:edot}-\ref{eq:gdot} were performed with the same parameters as in Fig.~\ref{fig:simulations} but with a spectral filter applied to the emitted photon.  To keep the results experimentally relevant, only a simple Lorentzian filter was used, representing a Fabry-Perot cavity, centered on the Raman-scattering frequency with width fixed at $20\xi$.  The results are shown in Fig.~\ref{fig:filter}(a), and indeed the degradation of photon indistinguishability matches Eq.~\ref{eq:indist_analytical_filter}.  Furthermore, there is little cost in terms of photon generation efficiency per cycle.  Even for the simple Lorentzian filter used in the simulation, the filtering loss was only $5\%$, and the theoretical photon collection efficiency decreased from $90\%$ to $85\%$.  In theory, if the ground states are completely free from dephasing processes then the photon indistinguishability can be improved to an arbitrary degree by decreasing the photon emission rate $\xi$, once $\xi < \beta$.  The main penalty is the decrease of $\xi$ itself, since this is the photon emission rate.  Decreasing $\xi$ will eventually reduce the communication or computation speed in any quantum network application.

The cavity can also serve as a spectral filter.  Until now, our calculations have been performed in a regime $\kappa \gg \Delta$ where the cavity served only to increase the effective spontaneous emission rate.  If instead we have $\kappa \ll \Delta$, the cavity can reject spontaneous emission noise at the natural transition frequency without the need for an external filter.  In the case of finite $\kappa$ the spectral transfer function $H(\omega,\xi)$ has three poles that contribute.  Combining Eq.~\ref{eq:g_approx}-\ref{eq:phi_dynamics} with Eq.~\ref{eq:Hdef} in the limits $\xi \ll \{\gamma,\gamma_p\} \ll \Delta$ with constant $\Omega$, we find the approximate expression,
\be
\fl
H(\omega, \eta) \approx \left|
\frac{\xi}{\gamma_p (i\omega-\eta/2)}
+ \frac{i\kappa}{(2\Delta+i\kappa)(i\omega-i\Delta-\gamma_{\mathrm{eff}}/2)}
- \frac{i\kappa}{(2\Delta+i\kappa)(i\omega-\kappa/2)} \right|^2 \, ,
\label{eq:H_threepoles}
\ee
where $\gamma_{\mathrm{eff}} = (\kappa^2 \gamma_p + 4\Delta^2 \gamma) / (\kappa^2 + 4\Delta^2)$ is the modified spontaneous emission rate from level $r$ including the cavity detuning.  Neglecting the cross-terms in Eq.~\ref{eq:H_threepoles}, the indistinguishability calculated using Eq.~\ref{eq:F_approx_freqdomain} becomes,
\be
\fl
1-F \approx
\frac{2\sigma^2}{\gamma_p^2(1+\frac{\beta}{\xi})} +
\frac{2 \sigma^2}{\Delta^2 + \beta^2} \frac{\gamma_{\mathrm{eff}} + 2\beta}{\gamma_{\mathrm{eff}}} \frac{\kappa^2}{\kappa^2 + 4\Delta^2}
 + \frac{8 \sigma^2 \kappa}{(\kappa^2+4\Delta^2)(\kappa+2\beta)} \, .
\label{eq:indist_goodcavity}
\ee
Comparing this with Eq.~\ref{eq:indist_analytical}, we see that the second term which corresponds to spontaneous emission noise at $\omega = \Delta$ is now suppressed by a factor $\kappa^2/ (\kappa^2 + 4\Delta^2)$ due to the cavity filtering.  The new, third term corresponds physically to resonant enhancement of noise by the cavity, and even for large $\Delta$ it can make a significant contribution.  To test the validity of Eq.~\ref{eq:indist_goodcavity}, Monte Carlo simulations of Eqs.~\ref{eq:edot}-\ref{eq:gdot} were performed for cases with $\kappa < \Delta$, and the results are shown in Fig.~\ref{fig:filter}(b) (see figure caption for parameters).  The results show that the cavity can indeed serve as an effective filter, and for $\beta \gg \xi$, photon indistinguishability far exceeding that of an equivalent two-level system is again obtained.

\section{Other Decoherence Mechanisms}

To estimate the {\it total} indistinguishability degradation in the Raman scheme we must include two other decoherence mechanisms in addition to pure dephasing of the excited-state.  The first is ground-state dephasing, against which the Raman scheme provides no protection. The photon indistinguishability degradation due to ground-state dephasing is similar to that in a two-level system,
\be
1 - F^\prime \approx \frac{2\sigma^{\prime 2}}{\xi(\xi+\beta^{\prime})} \, , \label{eq:ground_dephasing}
\ee
where $\sigma^{\prime}$ and $\beta^{\prime}$ are the ground-state fluctuation amplitude and rate, respectively, under the noise model of Eq.~\ref{eq:noisedef}.  In contrast with Eq.~\ref{eq:indist_goodcavity}, the contribution from ground-state dephasing increases as the photon emission rate $\xi$ is decreased.  In this model the effective decay rate $1/T_2^{\ast}$ of the ground-state coherence is $\sigma^{\prime 2}/\beta^\prime$ for large times ($t \gg 1/\beta$), but for short times the coherence decay follows a Gaussian function with characteristic timescale $\sim 1/\sigma^\prime$.

The second additional decoherence mechanism is spontaneous decay from $\ket{r}$ back to $\ket{e}$.  As discussed in Ref.~\cite{ref:kiraz2004qss}, this can be viewed as a time-jitter process, with the last $r \rightarrow e$ photon marking a random delay before the start of the emitted $r \rightarrow g$ photon.  The indistinguishability degradation resulting from this process is,
\be
1 - F^{\prime\prime} \approx \frac{\xi_e}{\xi_e + \xi} \, , \label{eq:pump_scattering}
\ee
where $\xi_e \approx \frac{|\W|^2}{4\D^2} \gamma_e$ is the decay rate back to $\ket{e}$ ($\gamma_e$ is the natural $r\rightarrow e$ spontaneous decay rate), and $\xi$ is the rate for the forward process as defined above.  In principle, for a three-level system $\xi_e$ can be made negligibly small by choosing system parameters such that the $e-r$ transition is very weak, and then compensating with a stronger excitation field.  In actual systems, the achievable improvement will be limited by additional excited states which, although they may be further detuned, will make a substantial contribution to the system dynamics when the excitation field is too strong.  The other way to improve the $\xi_e/\xi$ ratio is through selective cavity enhancement of the $g-r$ transition.  This requires either a cavity linewidth that is narrow compared with the $e-g$ energy spacing or else good polarization selection rules.

\section{Solid-State Sources of Indistinguishable Photons}

Let us now discuss how the above results may apply to actual solid-state atom-like systems that have two or more ground-state spin sublevels connected by {\it strong} optical transitions to a common excited state.  Various types of optical control have been demonstrated in semiconductor quantum dots~\cite{ref:dutt2005sso,ref:xu2007fss,ref:xu2008coherent,ref:press2008cqc,ref:kim2008osi,ref:vamivakas2009srq,ref:brunner2009csh,ref:fernandez2009ots}, shallow donors~\cite{ref:fu2008ucd}, and diamond NV centers~\cite{ref:hemmer2001res,ref:santori2006cpta,ref:santori2006cptb}.  The benefit provided by a large detuning will first of all depend on the fluctuation rate of the noise source affecting the excited state.  For spectral-diffusion processes with long correlation timescales we expect little improvement compared with a two-level scheme for useful photon emission rates, but a large improvement may be possible for noise processes that have sub-nanosecond correlation timescales.

First, let us consider a charged quantum dot subject to a strong magnetic field that is not aligned to the growth axis, so that two long-lived electron spin ground states are coupled to an excited (trion) state by optical transitions. Suppose this quantum dot is placed in a state-of-the-art photonic-crystal microcavity~\cite{ref:englund2007ccr} with cavity-QED parameters  $g = 2\pi \times 8 \, \mathrm{GHz}$ and $\kappa = 2\pi \times 33 \, \mathrm{GHz}$.  Because spontaneous emission into leaky modes in 2D photonic crystals is suppressed by a factor of $2-5$~\cite{ref:englund2005cse,ref:kress2005mse} we set $\gamma = (0.5 \, \mathrm{ns})^{-1}$.  Suppose now that the excited state is subject to a phonon dephasing process which we shall approximate using the above dephasing model with $1/T_{2,ex} = \sigma^2/\beta = 2\pi \times 0.16 \, \mathrm{GHz}$, with fluctuation rate $\beta = 2\pi \times 80 \, \mathrm{GHz}$ (corresponding to $kT$ at $T=4\,\mathrm{K}$).  If the system is excited on resonance with a short pulse, the best-case indistinguishability degradation due to excited-state dephasing is $1-F \approx 0.04$ with a photon temporal width (FWHM) of $\sim 35 \, \mathrm{ps}$ (the decay is non-exponential since the system is near the onset of strong coupling).  For non-resonant excitation, possibly the only option with photonic crystal cavities due to pump laser scatter, the indistinguishability will be much worse due to the time jitter associated with the finite relaxation time to the upper level of the optical transition.  Suppose now that we instead use a Raman scheme with $\Omega = 2\pi \times 16 \, \mathrm{GHz}$ and $\Delta = 2\pi \times 40 \, \mathrm{GHz}$, and send the collected light through a Fabry-Perot cavity with a transmission bandwidth of $2\pi \times 4 \mathrm{GHz}$.  The simulated filter efficiency is $88 \%$, and the total photon collection efficiency is $87 \%$.  The theoretical indistinguishability degradation in the Raman scheme, considering excited-state dephasing alone, is $1-F \approx 0.002$ while the lifetime for the Raman transition is $0.5 \, \mathrm{ns}$, still quite fast for many applications.

To predict the total degradation we must also consider the processes in Eqs.~\ref{eq:ground_dephasing},\ref{eq:pump_scattering}.  In negatively charged InAs quantum dots $1/T_2^\star$ is limited to $< 10 \, \mathrm{ns}$  due to hyperfine coupling between the electronic and nuclear spins.  However, a recent experiment with positively charged quantum dots~\cite{ref:brunner2009csh} suggests that $T_2^{\star}$ values of hundreds of nanoseconds or more may be achievable.  If we take $T_2^\star$ to be $500\,\mathrm{ns}$, for $\beta^\prime \gg \xi$ one could hope for $1-F^\prime = 2 / (\xi T_2^\star) = 0.002$, which approximately matches the degradation predicted above for excited-state dephasing.  If $\beta^\prime \ll \xi$ the situation is better since the dephasing will follow a Gaussian time dependence, and in the first $0.5 \, \mathrm{ns}$ may be negligible (this is reflected in the $\xi^{-2}$ dependence in Eq.~\ref{eq:ground_dephasing} for $\beta^\prime \rightarrow 0$).  The main limiting factor is likely to be spontaneous emission back to $\ket{e}$.  If we have a perfectly balanced $\Lambda$ system, the degradation is given approximately by the inverse Purcell factor $\left(\frac{4 g_0^2}{\kappa \gamma} \right)^{-1}$ giving $1-F^{\prime\prime} \sim 0.01$.  It is tempting to try to reduce $\gamma_e$ by moving to an imbalanced $\Lambda$ system using a magnetic field that is nearly aligned to the growth axis, or by using a very large magnetic field.  However there are in fact two excited (trion) states, and if $\gamma_e$ decreases for one of them, it will increase for the other.  Furthermore, at practical magnetic field strengths the energy separation between the two trion states cannot be made much larger than the value of $\Delta$ used in this example.  Thus, $1-F_{\mathrm{total}} \sim 0.01$ is probably close to the best possible performance.  Nevertheless, we note several improvements in the Raman scheme compared with the two-level scheme.  First, the overall indistinguishability is substantially improved, especially if we consider realistic excitation schemes in the two-level case.  Additionally, the Raman scheme can tolerate much more high-frequency excited-state dephasing than was included here. Finally, the Raman scheme includes a long-lived matter qubit coupled to a single-photon emitter as needed for quantum networking applications. In this example we assumed that the ground-state splitting could be made large compared with the cavity linewidth $\kappa$.  If this cannot be achieved in a photonic-crystal cavity using practical magnetic field strengths, then it may be advantageous instead to use microdisk cavities, which tend to have higher quality factors, can also reach a strong coupling regime, and can be coupled efficiently to a tapered optical fiber~\cite{ref:srinivasan2007lno}.

Next, let us briefly discuss the case of diamond NV centers.  This system seems attractive for quantum computation since in isotopically purified diamond the spin coherence lifetime (as measured in a spin-echo experiment) can reach several milliseconds, and $T_2^\star$ (without spin echo) can be several microseconds or longer~\cite{ref:balasubramanian2009usc}.  A $\Lambda$-type system can be obtained either by using a magnetic field to mix two of the ground states~\cite{ref:hemmer2001res} or by using strain or an electric field to mix some of the excited states~\cite{ref:santori2006cpta,ref:santori2006cptb,ref:tamarat2008sfs}.  Furthermore, at low temperatures, lifetime-limited spectral linewidths of $13 \, \mathrm{MHz}$ have been observed in high-purity samples using laser spectroscopy~\cite{ref:tamarat2006ssc}.  However, while much recent progress has been made towards the realization of high-$Q$, small-mode-volume cavities in diamond~\cite{ref:larsson2009com,ref:barclay2009cbm}, it seems unrealistic to expect extremely large factors for total spontaneous emission lifetime modification as have been realized in quantum dots.  Apart from technological issues, the main difficulty in this system is that only $\sim 4 \%$ of the spontaneous emission is into the zero-phonon line, and only this emission can be efficiently coupled to a high-$Q$ cavity.  This amounts to a factor of $25$ penalty in converting the Purcell factor into total spontaneous emission lifetime modification.  Thus, the best approach to obtain a photon indistinguishability above $0.9$ in a Raman scheme is probably to use an imbalanced $\Lambda$ system to suppress the $r \rightarrow e$ process.  Since the excited-state level splittings near an anticrossing are only $\sim 200 \, \mathrm{MHz}$~\cite{ref:tamarat2008sfs}, there is not much room for large detunings.  Furthermore, the most problematic excited-state dephasing process in this system seems to be a slow spectral diffusion process related to charge trap fluctuations as was mentioned above.  For these reasons, and because the spontaneous emission lifetime ($12 \, \mathrm{ns}$) is already rather long, the best performance can probably be obtained by operating the system on resonance.

\section{Conclusion}

In summary, we have shown that if photon indistinguishability is limited by excited-state dephasing, it can in some cases be drastically improved using a three-level Raman scheme with a large detuning.  The dependence of the photon indistinguishability on the noise fluctuation rate was examined, and it was shown that a large detuning provides an improvement when the noise fluctuation rate is fast compared with the photon emission rate.  To realize this benefit it is usually necessary to use an internal or external spectral filter to remove a weak spontaneous emission component at the natural transition frequency.  We have also considered the practical application of this scheme to solid-state systems such as semiconductor quantum dots and diamond nitrogen-vacancy centers.  Due to additional mechanisms which degrade the photon indistinguishability, a large detuning appears most helpful in the quantum-dot case where large Purcell enhancements can be achieved.  With this approach, we are optimistic that solid-state single-photon sources can be improved to a level where they will be truly useful for quantum network applications such as creating distributed entanglement between matter qubits.

This material is based upon work supported by the Defense Advanced Research Projects Agency under Award No. HR0011-09-1-0006 and The Regents of the University of California.

\end{document}